\documentclass[letterpaper,prd,superscriptaddress,twocolumn,aps,reprint,notitlepage]{revtex4-1}
\usepackage{amssymb,amsmath}
\usepackage[dvips]{graphicx}
\usepackage[usenames]{color}

\begin{document}

\title{Negative optical inertia for enhancing the sensitivity of future gravitational-wave detectors}

\date{\today}

\author{Farid Khalili}
\affiliation{Faculty of Physics, Moscow State University, Moscow 119991, Russia}
\author{Stefan Danilishin}
\affiliation{Faculty of  Physics, Moscow State University, Moscow 119991, Russia}
\author{Helge M\"uller-Ebhardt}
\affiliation{Max-Planck Institut f\"ur Gravitationsphysik (Albert-Einstein-Institut) and Leibniz Universit\"at Hannover, Callinstr. 38, 30167 Hannover, Germany}
\author{Haixing Miao}
\affiliation{School of Physics, University of Western Australia, WA 6009, Australia}
\author{Yanbei Chen}
\affiliation{Theoretical Astrophysics 130-33, California Institute of Technology, Pasadena, CA 91125, USA}
\author{Chunnong Zhao}
\affiliation{School of Physics, University of Western Australia, WA 6009, Australia}

\begin{abstract}

We consider enhancing the sensitivity of future gravitational-wave detectors by using double optical spring.  When the power, detuning and bandwidth of the two carriers are chosen appropriately, the effect of the double optical spring can be described as a ``negative inertia", which cancels the positive inertia of the test masses and thus increases their response to gravitational waves. This allows us to surpass the free-mass Standard Quantum Limit (SQL) over a broad frequency band, through signal amplification, rather than noise cancelation, which has been the case for all broadband SQL-beating schemes so far considered for gravitational-wave detectors.  The merit of such signal amplification schemes lies in the fact that they are less susceptible to optical losses than noise cancelation schemes.  We show that it is feasible to demonstrate such an effect with the {\it Gingin High Optical Power Test Facility}, and it can eventually be implemented in future advanced GW detectors.
\end{abstract}

\maketitle

\section{Introduction}

The Heisenberg Uncertainty Principle, when applied to test masses, has long been known to impose a so-called Standard Quantum Limit (SQL) for high-precision displacement and force measurements\,\cite{67a1eBr, 92BookBrKh}. In essence, SQL corresponds to the point where the measurement noise, which is inversely proportional to the coupling strength between the meter and the test object, becomes equal to the back-action noise, which arises from the test object perturbation by the meter, and is directly proportional to the coupling strength.

Contemporary first-generation large-scale laser interferometric gravitational-wave (GW) detectors (LIGO\,\cite{Abramovici1992, LIGOsite}, Virgo\,\cite{Ando2001, VIRGOsite}, GEO600\,\cite{Willke2002,GEOsite}, and TAMA\,\cite{TAMAsite}) have not yet reached this limit. In these devices, the measurement noise results from fundamental quantum fluctuation in the phase of the light, which is also called the shot noise; the back action noise arises from quantum fluctuation in the amplitude of the light\,\cite{Caves1981}, which exerts a random radiation pressure force on the test object, and thus also called the radiation-pressure noise. The measurement sensitivity is determined by the amount of optical power circulating in the interferometers. For the first generation GW detectors, this is quite high, up to tens of kilowatts, but it is still insufficient to ``feel'' the quantum radiation pressure noise. Second-generation detectors, e.g., Advanced LIGO, Advanced Virgo, GEO-HF and LCGT, aim at increasing sensitivity by about one order of magnitude by increasing optical power, improving the optics, and evolutionary changes of the interferometer configurations\,\cite{Thorne2000, Fritschel2002, Smith2009, Acernese2006-2, Willke2006, LCGTsite}. As a result, it is anticipated that the second-generation detectors will be \textit{quantum noise limited}: at high frequencies, the main sensitivity limitation will be due to the shot noise, and at low frequencies, due to the radiation pressure noise. At the point of the best sensitivity, where these two noise strengths become equal the SQL will be reached.

To overcome such a quantum barrier that limits the sensitivity for detecting GWs, several approaches have been proposed. They fall into two main categories: the first one comprises {\it noise-cancelation} schemes. It uses the fact that the goal of GW detectors is not the measurement of the test masses position, which is a {\it quantum} variable and thus can not be measured continuously with precision better that the SQL, but rather the detection of GW signals, which can be treated as {\it classical} forces acting on the test masses \cite{03a1BrGoKhMaThVy}. It was shown in Ref.\,\cite{Unruh1982} that, by introducing cross-correlation between the measurement noise and the back action noise, the latter one can be canceled, and thus in principle arbitrarily high sensitivity can be achieved. Realistic topologies based on this principle, which probably will be implemented in the third-generation GW detectors, was proposed\,\cite{02a1KiLeMaThVy, 00a1BrGoKhTh, Purdue2002, Chen2002}. Unfortunately, the inherent disadvantage of such schemes is that they are very sensitive to optical losses (in particular, the non-unity quantum efficiency of the photodetector, which destroys quantum correlations).  A rule of thumb for the limit of  achievable SQL-beating in this case can be written as (refer to Ref.\,\cite{09a1ChDaKhMu} for more details):
\begin{equation}
\xi =\sqrt{{S_h}/{S_h^{\rm SQL}}} \gtrsim ( e^{-2q} \epsilon)^{1/4}\,.
\end{equation}
Here $S_h$ is the noise spectral density of the detector in terms of GW strain $h$, and $S_h^{\rm SQL}$ is the corresponding SQL; $\epsilon$ quantifies the optical loss, and $e^{-2q}$ is the squeezing factor if nonclassical squeezed light is implemented.  Even for rather optimistic values of the optical parameters with $\epsilon=0.01$ and $e^{-2q}=0.1$ (10~dB squeezing), we have $\xi\gtrsim 5.6$, which means that one can only surpass the SQL by approximately a factor of five with the noise-cancelation schemes.

The second group of methods is based on amplification of the detector response to the GW signal by modifying the test-mass dynamics, which we can call the {\it signal amplification} schemes. This is based on the fact that the SQL for a force (e.g., GW tidal force in our case) measurement $S_F^{\rm SQL}$ depends on the test-mass dynamics, which, more explicitly, can be written as
\begin{equation}\label{S_F_SQL_gen}
S_F^{\rm SQL}(\Omega) = 2\hbar|\chi^{-1}(\Omega)| \,,
\end{equation}
where $\Omega$ is the frequency of the signal, and $\chi$ is the mechanical susceptibility of the test mass, which is the ratio of  the test-mass displacement $x(\Omega)$ to the acting force $F(\Omega)$: $ \chi (\Omega) =x(\Omega)/F(\Omega).$ For a mechanical probe body, $\chi(\Omega) = [m(\omega_m^2-\Omega^2)]^{-1}$ with $m$ and $\omega_m$ being its mass and mechanical eigenfrequency, respectively. It has a much stronger response to near-resonance force, and thus a smaller SQL.  In a typical terrestrial GW detector, the characteristic eigenfrequency (pendulum mode) of the test mass is around 1~Hz, and this is much smaller than that of the GW signals around 100~Hz, and the corresponding SQL is almost identical to that of a free mass:
\begin{equation}
[S_F^{\rm SQL}]_{\rm free\,mass}=2\hbar\, m\Omega^2.
\end{equation}
To surpass this SQL, a natural idea is to modify the test-mass dynamics, and to upshift the eigenfrequency to be near 100 Hz. Such a sensitivity improvement is obtained not by a delicate cancelation of quantum noise, but by amplification of the signal, thus much less susceptible to the optical losses. Recently, this sensitivity gain was experimentally demonstrated using very small mechanical oscillator (nano-beam) with microwave position sensor\,\cite{Teufel2009}. Near the mechanical resonance frequency (about 1\,MHz), the achieved sensitivity was several times better than  the free-mass SQL, albeit much worse than the harmonic oscillator SQL at this frequency (it was limited by the nano-beam thermal noise).

\begin{figure}
\includegraphics[width=0.35\textwidth, bb=0 0 518 108, clip]{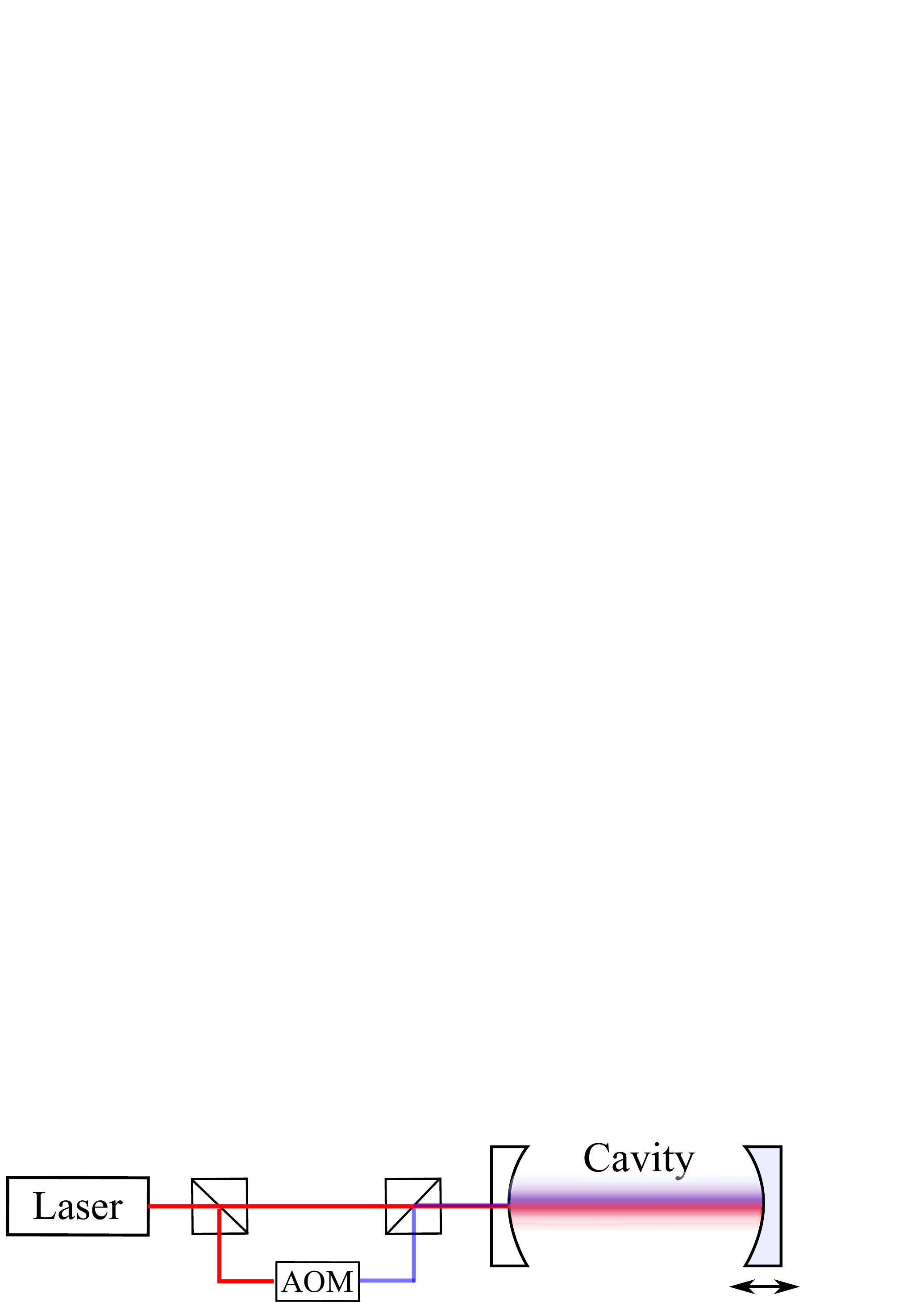}
\caption{A schematic plot showing the experimental realization of double optical spring, as demonstrated experimentally\,\cite{Corbitt2007}. The other carrier light is obtained by shifting the laser frequency with an acoustic-opto-modulator (AOM). \label{scheme}}
\end{figure}

In GW detectors, ordinary mechanical oscillators of solid-state springs cannot be used due to unacceptable high technical noise, and also unattainable high stiffness of the material (a km-scale spring of 100 Hz frequency). To overcome this difficulty, a  low-noise {\it optical spring}, which arises in detuned Fabry P\'{e}rot cavities,  can be used instead\,\cite{99a1BrKh, Buonanno2001, Buonanno2002}. With a high optical power circulating inside the cavity, the radiation pressure force on the test mass varies dramatically as the test-mass position changes, which effectively creates a highly-rigid spring. The test-mass dynamics (more specifically that of the differential motion of the input and end test masses in the arms of a Fabry-P\'{e}rot--Michelson GW interferometer)  will be modified as:
\begin{equation}
-m\Omega^2 x(\Omega)= - K(\Omega) x(\Omega)+ F(\Omega)\,,
\end{equation}
where we have ignored the low pendulum frequency of the test mass, and $K(\Omega)$ is the optical rigidity. The resulting modified mechanical susceptibility reads
\begin{equation}
\chi(\Omega) =[{ -m\Omega^2 +K(\Omega)}]^{-1}.
\end{equation}
The sign of optical rigidity depends on the sign of the cavity detuning (the difference between the laser frequency $\omega_0$ and the cavity resonant frequency $\omega_c$). A blue-detuned pumping ($\omega_0> \omega_c$) creates a positive rigidity, while a red-detuned one ($\omega_0< \omega_c$) creates a negative rigidity. In addition, the rigidity is accompanied by a damping of the opposite sign: a positive rigidity with a negative damping, and vice versa, and, therefore, a single optical spring is always unstable. Recently, it has been shown theoretically, and demonstrated experimentally\,\cite{Corbitt2007} that a stable configuration, with both positive rigidity and positive damping, can be obtained by pumping the cavity with lasers at different frequencies (one blue-detuned and the other red-detuned with respect to the cavity eigenfrequency), of which the experimental setup is shown schematically in Fig.\,\ref{scheme}. This combines  two optical springs of opposite signs, which is the so-called  {\it double optical spring}.

With the double optical spring, as shown in Ref.\,\cite{Rehbein2008}, the test-mass frequency can indeed be shifted up to 100~Hz in future advanced GW detectors. The optical rigidity $K(\Omega)$ can have sophisticated frequency dependence, and it allows us to overcome the shortcoming of ordinary oscillators --- a narrow frequency band enhancement--- and to achieve a broadband enhancement of the sensitivity. As we will see, by properly tuning the cavity, the optical rigidity can have the following frequency dependence:
\begin{equation}
K(\Omega) \approx - m_{\rm opt}\Omega^2\,,
\end{equation}
with $m_{\rm opt}$ a constant over a broad frequency band, acting as an additional {\it electromagnetic inertia}.  When this inertial $m_{\rm opt}$ is negative, we will have
\begin{equation}
\chi(\Omega) \approx \frac{1}{-(m+m_{\rm opt}) \Omega^2} ={\frac{1}{(|m_{\rm opt}|-m)\Omega^2}} \,,
\end{equation}
which is greatly enhanced compared to the free-mass susceptibility over a broad frequency band if $|m_{\rm opt}|\sim m$. It stands to mention that such a negative inertia has also been studied previously in Ref.\,\cite{HelgeThesis}, when considering Sagnac interferometers with detuned signal recycling. At low frequencies,  the out-going field is proportional to the speed of test-mass motion, and radiation-pressure force is in turn proportional to the time derivative of the in-going field fed back by the signal-recycling mirror, and hence two time derivatives are taken on the test-mass position, before it is re-applied as radiation-pressure force. The other point of view is that the signal recycling Sagnac has two effective optical resonators coupled to the test mass, playing the role of the two optical springs here.

This paper is organized as follows:  in Sec.\,\ref{sec:simple}, we will introduce the negative inertia effect, derive the necessary conditions to achieve the required frequency dependence, and estimate the enhancements allowed;  in Sec.\,\ref{sec:gingin}, we will consider a possible experimental demonstration of this effect using the {\it Gingin High Optical Power Test Facility}\,\cite{AIGOsite}; in  Sec.\,\ref{subsec:ligo}, we will consider its application to future large-scale gravitational-wave detectors. Finally, in Sec.\,\ref{sec:conclusion}, we will conclude our main results.

\section{Negative optical inertia}\label{sec:simple}

\subsection{Beating the free-mass SQL by modifying dynamics}
\label{sec:sql:overview}

\begin{figure}[!h]
\includegraphics[width=0.35\textwidth]{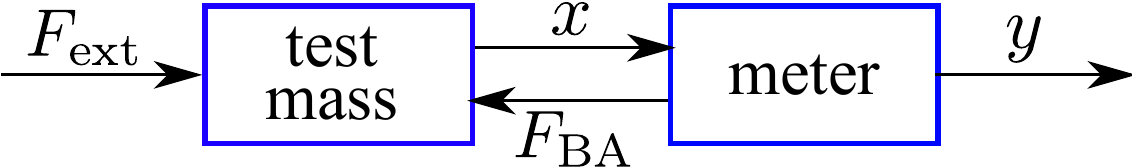}
\caption{A schematic plot showing a linear quantum measurement device. In the context of GW detection,
the external force $F_{\rm ext}$ is the GW tidal force. The meter is the optical field that measures the
test-mass position $x$, and at the same time, exerts a radiation pressure force (back action $F_{\rm BA}$).
\label{detection}}
\end{figure}

Before giving the details of the negative inertia idea, it is illuminating to first discuss the linear quantum
measurement, to see how the SQL is imposed and how the free-mass SQL can be surpassed by modifying dynamics.
In Fig.\,\ref{detection}, we show a typical linear measurement device which includes the GW detector as a special
case. The meter measures the displacement of the test mass to infer the external force that is acting on the test mass.
The dynamics of the system is governed by a set of linear equations, which, in
the frequency domain, reads
\begin{align}
 x(\Omega) &= \chi(\Omega)[F_{\rm BA}(\Omega)+F_{\rm ext}(\Omega)],\\
y(\Omega) &=Z(\Omega) + x(\Omega),
\end{align}
where $Z(\Omega)$ is the measurement noise of the meter output $y$.
The output can be decomposed into the signal part $y_s=\chi\, F_{\rm ext}$ and the noise part $y_n=Z+\chi\, F_{\rm BA}$,
of which the spectral density, normalized with respect to $F_{\rm ext}$, is
\begin{equation}
S^{\rm noise}_F(\Omega)={|\chi^{-2}(\Omega)|}{S_Z(\Omega)}+S_F(\Omega),
\end{equation}
where $\langle Z(\Omega)Z^{\dag}(\Omega')\rangle\equiv \pi\,S_Z(\Omega)\delta(\Omega-\Omega)$ and for the back action noise
$\langle F_{\rm BA}(\Omega)F_{\rm BA}^{\dag}(\Omega')\rangle\equiv \pi\,S_F(\Omega)\delta(\Omega-\Omega)$. We assume that $Z$ and $F_{\rm BA}$ are not correlated, \textit{i.e.} that $S_{ZF}(\Omega)=0$. For a quantum-limited
meter, the Heisenberg Uncertainty Principle imposes the following constraint on the spectral
density of $Z$ and $F_{\rm BA}$\,\cite{92BookBrKh}:
\begin{equation}
S_{Z}(\Omega) S_{F}(\Omega)\ge \hbar^2\,.
\end{equation}
The meter sensitivity then will be limited by the SQL in acordance with Eq.\,\eqref{S_F_SQL_gen}:
\begin{equation}
S_F^{\rm noise}(\Omega)\ge 2|\chi^{-1}(\Omega)|\sqrt{S_Z(\Omega)S_F(\Omega)}\ge 2\hbar |\chi^{-1}(\Omega)|.
\end{equation}
It is clear that, the higher the classical susceptibility $|\chi|$, the smaller is the SQL, and the better sensitivity an SQL-limited meter can achieve \textit{in principle}, with the only limitation coming from the classical force noise which enters in the same way as the signal.  By modifying the dynamics of the nearly free test masses with the negative inertia, we can decrease the SQL by the following factor:
\begin{equation}
{\frac{[S_{F}^{\rm SQL}]_{\rm modified}}{[S_F^{\rm SQL}]_{\rm free\,mass}}}=\left|\frac{[\chi(\Omega)]_{\rm modified}}{m\Omega^2}\right|={\left|\frac{|m_{\rm  opt}|-m}{m}\right|},
\end{equation}
which can be arbitrarily small if $|m_{\rm opt}|\rightarrow m$.
The advantage of modifying dynamics is that the signal is amplified at its origin, helping the
signal to pass through noisy channels and thus being more robust against optical losses than those noise-cancelation schemes.

However, in order to really follow the new SQL in a broad frequency band, we need to tailor the response of the SQL-limited meter, in such a way that
\begin{equation}\label{cond3}
S_{Z}(\Omega) = |\chi(\Omega)|^2 S_{F}(\Omega)\,.
\end{equation}
Therefore, non-trivial frequency dependence of the noise spectral densities, which follows frequency dependence of $\chi(\Omega)$, is required, which is clearly not always achievable. In the case of $\chi(\Omega)\propto 1/\Omega^{2}$ (a free-mass like response),  the sensing strategy that realizes such a requirement in a broad frequency band turns out to be speed measurement, which has shot noise $\propto1/\Omega^2$ and back-action noise $\propto\Omega^2$ --- speed meters can be realized by dual-cavity Michelson~\cite{00a1BrGoKhTh, Purdue2002} configurations and Sagnac interferometers~\cite{Chen2002}. In the later discussions, we will assume that
such a frequency dependence of the noise spectral densities can be satisfied.

\subsection{Negative Inertia: the idea}

In this section, we will discuss how to realize the negative inertia in details. To simplify the discussion, we will
use the conclusion in Ref.\,\cite{Buonanno2003} to map an interferometric GW detector into a
single detuned Fabry-P\'{e}rot optical cavity, with doubled circulating power $I_c$ and effective mass $m$.
We can therefore consider only a single cavity, of which results can be directly mapped back to the interferometer
case.

As shown in Ref.\,\cite{01a2Kh,Buonanno2003}, given a detuned Fabry-P\'erot cavity, the frequency-dependent optical
rigidity is equal to:
\begin{equation}\label{K}
K = \frac{mJ\delta}{ -\Omega^2 -2i\gamma\Omega + \Delta^2} \,.
\end{equation}
Here $\gamma$ is the cavity bandwidth; $\Delta^2=\delta^2+\gamma^2$ with $\delta=\omega_0-\omega_c$ being
the cavity detuning;
$J = {4\omega_0 I_c}/({mcL})$ with $I_c$ the optical power circulating inside the cavity and $L$ the cavity length.

The parameter regime, which we are concerned with, is that both $\Omega$ and $\gamma$ are small in comparison with the detuning $\delta$. Correspondingly, $K$ can be expanded in Taylor series over $\Omega$:
\begin{equation}
K \approx \bar K - i\,\Gamma_{\rm opt}\,\Omega  - {m}_{\rm opt}\Omega^2 + {\cal O}(\Omega^3) \,,
\end{equation}
Here
\begin{align}
& \bar K = \frac{mJ \delta}{\Delta^2} \,, & & \Gamma_{\rm opt}= -\frac{2mJ\gamma}{\Delta^4} \,, &
& m_{\rm opt} = -\frac{mJ\delta}{\Delta^4}
\end{align}
are the static rigidity, the optical damping and the {\it effective electromagnetic inertia} factor, respectively.  Note that, similar to $\bar K$  and $\Gamma_{\rm opt}$, the electromagnetic inertia $m_{\rm opt}$ can be either positive or negative, depending on the  sign of detuning $\delta$. It is therefore possible to combine two optical carriers with different powers, bandwidths and detunings, i.e., $(J_1,\gamma_1,\delta_1) \neq (J_2,\gamma_2,\delta_2)$, in such a way that their static rigidities $\bar K_{1}, \bar K_{2}$ cancel each other, and the total optical inertia cancels the mechanical inertia of the test mass, namely
\begin{equation}\label{cond2}
\bar K_{1} + \bar K_{2} = 0 \,,\quad m +m_{\rm opt1} + m_{\rm opt2}= 0\,.
\end{equation}
This will result in an effective test object that has high susceptibility, compared to a free mass, in a broad band.

More specifically, with double optical spring,  the mechanical susceptibility will be modified as
\begin{align}\label{rchi}
&\chi^{-1}(\Omega)= -m\Omega^2 + \frac{mJ_1\delta_1}{\mathcal{D}_1}
  + \frac{mJ_2\delta_2}{\mathcal{D}_2}  \nonumber \\
&= \frac{m}{\mathcal{D}_1\mathcal{D}_2}\Bigl[
    s^6 + 2(\gamma_1+\gamma_2)s^5 + (\Delta_1^2 + \Delta_2^2 + 4\gamma_1\gamma_2)s^4 \nonumber\\
  & \;\;\;+ 2(\gamma_1\Delta_2^2+\gamma_2\Delta_1^2)s^3 + (\Delta_1^2\Delta_2^2 + J_1\delta_1 + J_2\delta_2)s^2 \nonumber  \\
    &\;\;\;+ 2(J_2\delta_2\gamma_1 + J_1\delta_1\gamma_2)s + (J_1\delta_1\Delta_2^2 + J_2\delta_2\Delta_1^2)
  \Bigr] ,
\end{align}
where ${\cal D}_i\equiv s^2+2\gamma_i s + \Delta_i^2 $, and $s=-i\Omega$.
The physical conditions in Eq.\,\eqref{cond2} for cancelation of the total rigidity and inertia, mathematically, are equivalent to
making those terms proportional to $s^2$ and $s^0$ in the above equation vanish, namely
\begin{equation}
J_1\delta_1\Delta_2^2 + J_2\delta_2\Delta_1^2 = 0\,,\;
\Delta_1^2\Delta_2^2 + J_1\delta_1 + J_2\delta_2 = 0 \,.
\end{equation}
Here we have chosen to eliminate leading terms in the numerator, instead of making a Taylor expansion of the susceptibility at low frequencies and eliminate those leading terms, because the current approach makes the resulting dynamical system more easily treatable: zeros of $\chi^{-1}$, i.e., eigenfrequencies of the new dynamics, are more easily solvable from parameters of the optical system.  It can be demonstrated to give similar results to that of the Taylor expansion at low frequencies. These two conditions are easy to satisfy if $J_1$ and $J_2$ are:
\begin{align}\label{eqJ}
& J_1 = \frac{\Delta_1^4\Delta_2^2}{\delta_1(\Delta_2^2-\Delta_1^2)} \,, &
& J_2 = \frac{\Delta_1^2\Delta_2^4}{\delta_2(\Delta_1^2-\Delta_2^2)} \,.
\end{align}
In order to compensate the static rigidity, the detunings have to be opposite. Since $J_{1,2}$ are, by definition, positive quantities, the larger by absolute value detuning has to be negative. In the later discussions, we assume that $|\delta_1|<|\delta_2|$, $\delta_1>0$, and $\delta_2<0$.

Unfortunately, the resulting mechanical susceptibility given Eq.\,\eqref{eqJ} corresponds to a dynamically unstable system. For small values of $\gamma_{1,2}\ll\delta_1$, the characteristic instability time can be approximated as follows:
\begin{equation}
\tau_{\rm instab} \approx \left(
  2\,\frac{J_2\delta_2\gamma_1 + J_1\delta_1\gamma_2}{\Delta_1^2+\Delta_2^2}
\right)^{-1/3} .
\end{equation}
Note that it depends on the bandwidths only as $\gamma^{-1/3}$. Therefore, even for small $\gamma_{1,2}\ll\Omega$, the instability time can be well within the working frequency band, $\Omega\,\tau_{\rm instab}\sim1$. This problem can be solved in two ways. First, {\it partial} compensation of the mechanical inertia is possible:
\begin{align}\label{eqJ_alpha}
& J_1 = \frac{\alpha\Delta_1^4\Delta_2^2}{\delta_1(\Delta_2^2-\Delta_1^2)} \,, &
& J_2 = \frac{\alpha\Delta_1^2\Delta_2^4}{\delta_2(\Delta_1^2-\Delta_2^2)} \,,
\end{align}
where $0<\alpha<1$ is the compensation factor. The remaining non-zero inertia $(1-\alpha)m$ stabilizes the system, giving the following instability time:
\begin{equation}
\tau_{\rm instab} \approx \left(
  \frac{2\alpha}{1-\alpha}\,
  \frac{J_2\delta_2\gamma_1 + J_1\delta_1\gamma_2}{\Delta_1^2\Delta_2^2}
\right)^{-1}.
\end{equation}
In this case, $\tau_{\rm instab}\sim\gamma^{-1}\gg\Omega^{-1}$, and the instability can be damped by an out-of-band feedback system.

The second way is to cancel, in addition to the rigidity and inertia, also the friction (the term proportional to $s$ in Eq.\,\eqref{rchi}.
It can be achieved by adjusting the bandwidths $\gamma_{1,2}$ in the following way:
\begin{equation}\label{gammas_ratio}
\frac{\gamma_2}{\gamma_1} = -\frac{J_2\delta_2}{J_1\delta_1} = \frac{\Delta_2^2}{\Delta_1^2},
\end{equation}
which, experimentally, can be realized by using the signal-recycling configuration\,\cite{Mizuno1993, Buonanno2003}. It can be shown that the remaining transfer function is equal to:
\begin{align}\label{rchi3}
\chi^{-1}(\Omega) = \frac{m}{\mathcal{D}_1\mathcal{D}_2}&\Bigl[
  s^6 + 2(\gamma_1+\gamma_2)s^5
  +(\Delta_1^2 + \Delta_2^2 + 4\gamma_1\gamma_2)s^4\nonumber\\&+2(\gamma_1\Delta_2^2+\gamma_2\Delta_1^2)s^3
\Bigr],
\end{align}
which corresponds to a dynamically-stable system, as can be easily shown by the Routh-Hurwitz criterion.
In addition, this system is very responsive. Keeping the leading in $\Omega$ (the cubic one) term in Eq.\,\eqref{rchi3},
the SQL, with this new modified dynamics, can be approximated as follows [cf. Eq.\,\eqref{S_F_SQL_gen}]:
\begin{equation}
[S_F^{\rm SQL}]_{\rm new}
  \approx 4\hbar m\left(\frac{\gamma_1}{\Delta_1^2} + \frac{\gamma_2}{\Delta_2^2}\right)\Omega^3
  \,,
\end{equation}
which corresponds to the following sensitivity gain, in comparison with a free mass:
\begin{equation}\label{SQL_gain}
\frac{[S_F^{\rm SQL}]_{\rm new}}{[S_F^{\rm SQL}]_{\rm free\,mass}}=\frac{[S_h^{\rm SQL}]_{\rm new}}{[S_h^{\rm SQL}]_{\rm free\,mass}} \approx \frac{\gamma\Omega}{\delta^2} \,,
\end{equation}
where, in the first equality, we have converted the force spectral density to that referred to the GW strain (the usual way of measuring sensitivity in GW detection).

\subsection{Potential Gain in Sensitivity}

As discussed in Sec.~\ref{sec:sql:overview}, the gain in sensitivity mentioned above has to be considered only as a {\it potential} one. To have the sensitivity at the level of the new SQL over a broad band,  we need some optimally tuned measuring device (for example, an additional third optical pumping), attached to this high-susceptibility test object. The corresponding
measurement noise and the back action noise of this meter should satisfy Eq.\,\eqref{cond3}.
If they are also uncorrelated with $S_{ZF}=0$ and Heisenberg limited with
$S_Z S_F = \hbar^2$, the spectral density of the measurement noise is therefore given by:
\begin{equation}
S_Z = \hbar |\chi(\Omega)| \propto \frac{\hbar \Delta_1^2\Delta_2^2}{2m(\gamma_1\Delta_2^2+\gamma_2\Delta_1^2)\Omega^3}.
\end{equation}

Concerning the back action part $S_F$, there are additional contributions from the radiation-pressure noises of the two carriers which create the double optical spring. Spectral densities of these noise sources are equal to (cf.~\cite{Buonanno2003}):
\begin{equation}
S_{F_{1,2}}= \frac{2\hbar mJ_{1,2}\gamma_{1,2}(\Delta_{1,2}^2+\Omega^2)}
  {|\mathcal{D}_{1,2}|^2} \,.
\end{equation}
However, only a fraction of the above noise affects the sensitivity irretrievably. This is because the cavity bandwidths $\gamma_{1,2}$, which appear in the numerator, each consist of two parts: (i) the one owing to transmissivity of the mirrors and (ii) the one, resulting from the optical losses (absorption and scattering). The information loss due to the mirrors transmissivity can be recovered by a means of additional photodetectors which output can be used to recover the outgoing information about the back-action-induced motion of the test masses. It is this optical loss dominated, irretrievable part of the radiation-pressure noise of each carrier that eventually degrades the sensitivity, and the sum spectral density of this additional noise is given by:
\begin{equation}\label{S_BA}
S_F^{\rm add} = 2\hbar m\gamma_{\rm loss}\left(
  \frac{J_1(\Delta_1^2+\Omega^2)}{|\mathcal{D}_1|^2}
  + \frac{J_2(\Delta_2^2+\Omega^2)}{|\mathcal{D}_2|^2}
\right) ,
\end{equation}
where $\gamma_{\rm loss} = {cA^2}/({4L})$
and $A^2$ is the optical loss per bounce in the cavity. This spectral density corresponds to the following sensitivity degradation, in comparison with the free-mass SQL: ${S_F^{\rm add}}/({2\hbar m\Omega^2}) \sim {\gamma_{\rm loss}\delta}/{\Omega^2}$.

We leave the question of how exactly one can achieve optical loss-limited sensitivity in a real GW interferometer open and will address it in our follow-up paper \cite{Vo_DOS}.
\begin{figure*}[t]
\includegraphics[width=0.5\textwidth]{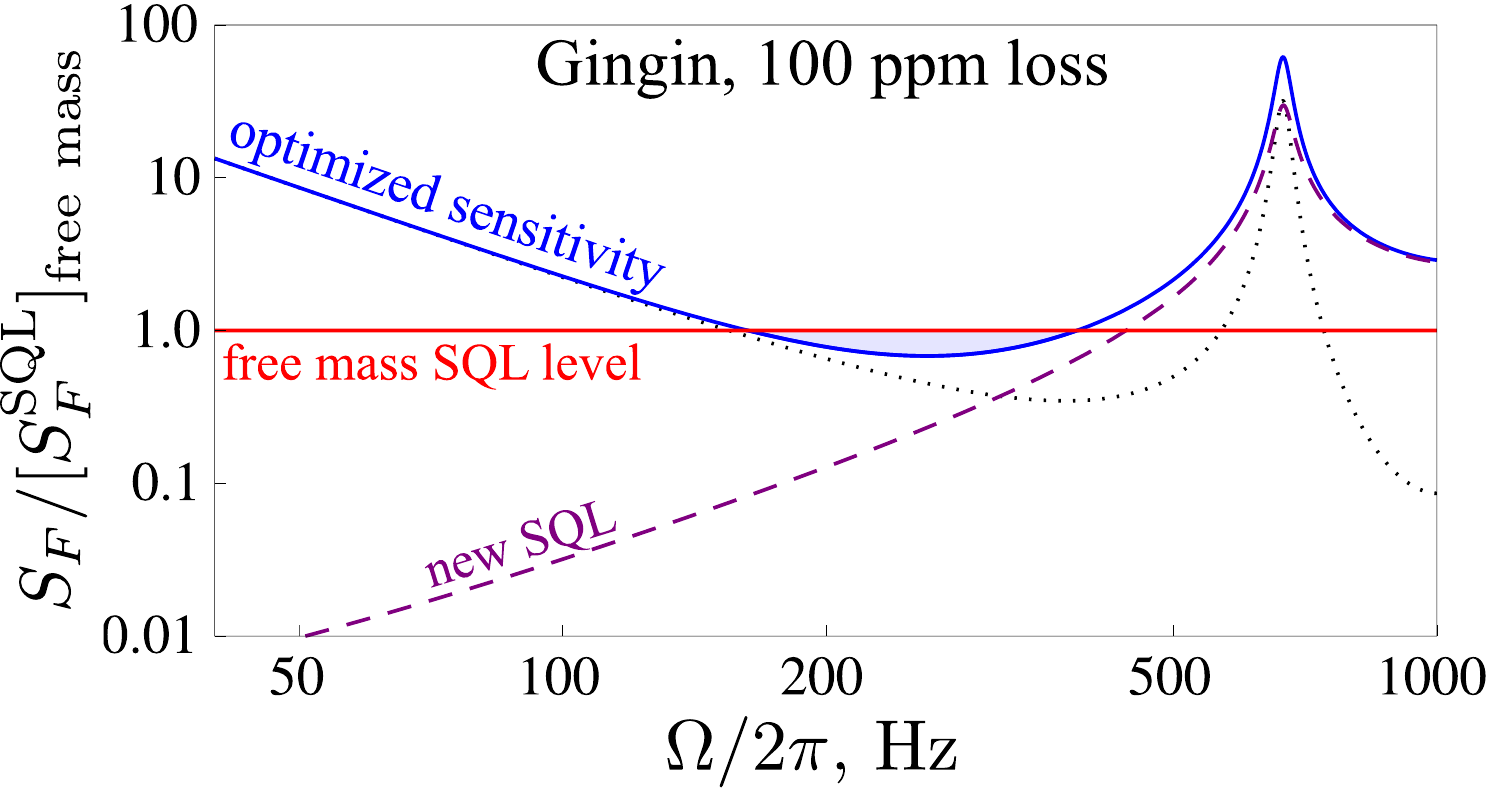}
\includegraphics[width=0.465\textwidth]{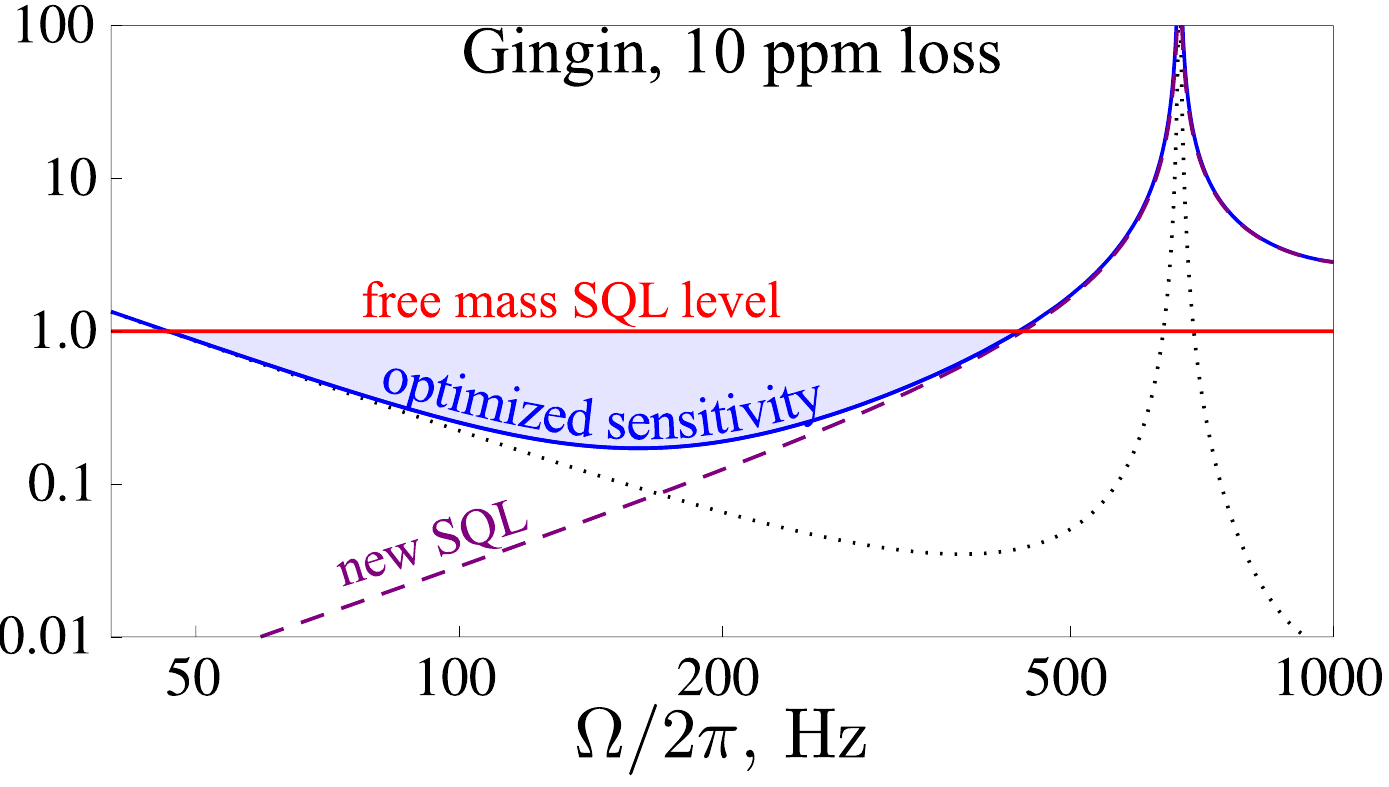}
\includegraphics[width=0.5\textwidth]{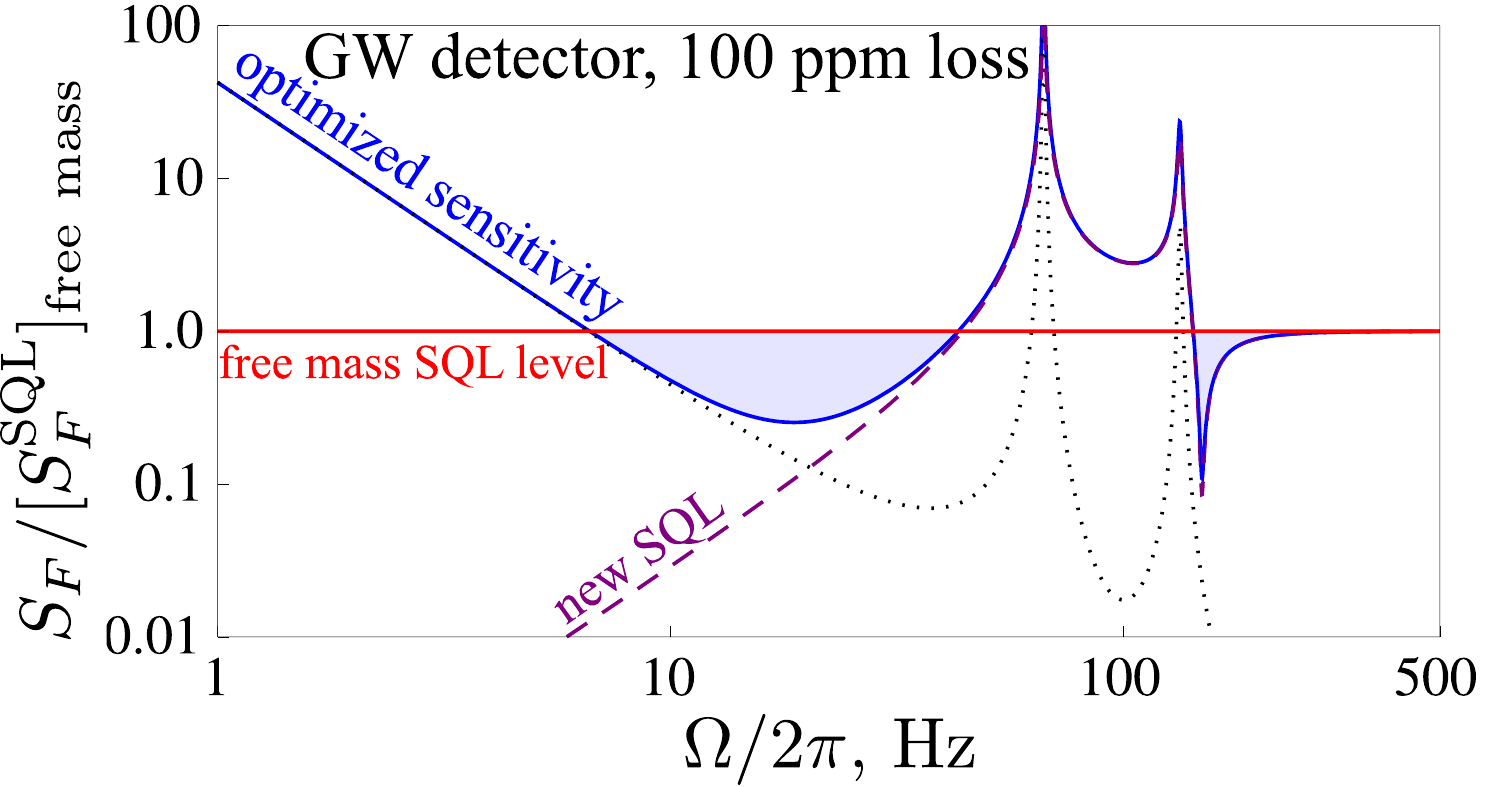}
\includegraphics[width=0.465\textwidth]{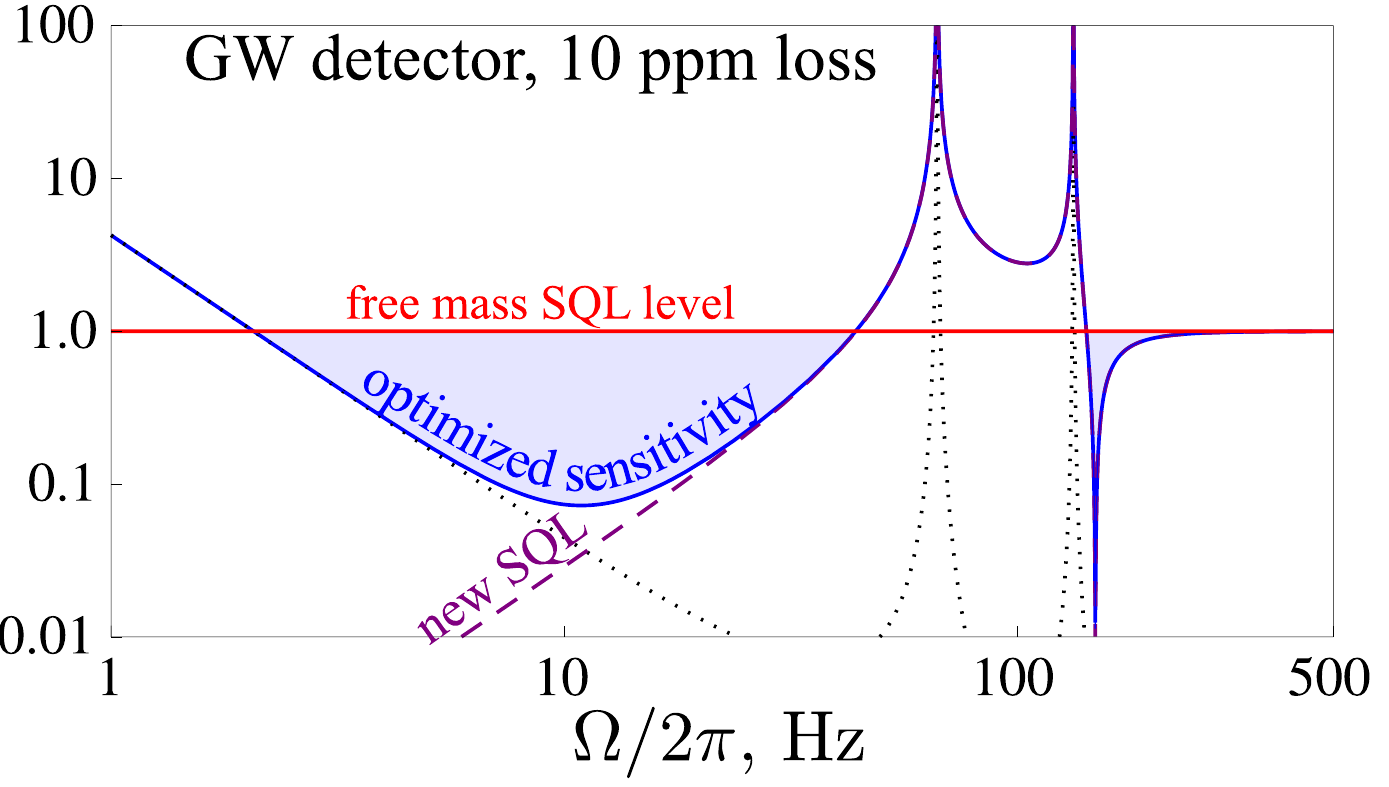}
\caption{Plots, showing the sum optimized noise spectral density $S_F^{\rm opt}=2\hbar |\chi^{-1}|+S_F^{\rm add}$ (blue solid line) normalized with respect to that of the free-mass SQL $[S_F^{\rm SQL}]_{\rm free\ mass}=2\hbar m\Omega^2$. The shaded area shows where the sensitivity surpasses the free-mass SQL. The dashed line shows the SQL $2\hbar |\chi^{-1}|$ with the new dynamics, and dotted line shows the spectral density $S_F^{\rm add}$ of the additional back-action noise due to optical loss.
The top row uses the specifications that are close to those of the Gingin facility with intracavity power of 100 kW.  The bottom row is similar to the AdvLIGO specifications: $L=4$ km, $m=40\,{\rm kg}$ and a total intracavity power of 2 MW for two carriers. Left column: optical losses per bounce $A^2=10^{-4}$, right column: $A^2=10^{-5}$.}\label{fig:plots}
\end{figure*}

\section{Experimental Realizations}

\subsection{The Gingin High Optical Power Test Facility}
\label{sec:gingin}

It follows from the above consideration, that the working frequency band of the negative inertia system is limited by $\gamma_{\rm loss}$ from below and by the detunings $\delta_{1,2}$ from above. The detunings, in turn, are limited by available optical power: it follows from Eqs.\,\eqref{eqJ}, that $\delta\sim J^{1/3}$. Therefore, the experimental demonstration of the negative optical inertia effect requires high-power interferometer with high-reflectivity mirrors and long arm(s) length. Among the prototype interferometers available or planned now, the {\it Gingin High Optical Power Test Facility }\,\cite{AIGOsite} is a good candidate for demonstrating this experiment.
The facility consists of a prototype interferometer with two 80 meter long optical cavities and $0.1$ kg test masses. Given future 50W laser input, the intracavity power can build up to 100 kW in the high-finesse cavity with an optical loss around 100 ppm ($A^2=10^{-4}$).

For a numerical estimate, we assume that the smaller bandwidth $\gamma_1$ is determined by the optical losses:
\begin{equation}
\gamma_1 = \gamma_{\rm loss} \approx 10^2\,{\rm s}^{-1} \,.
\end{equation}
In order to determine the other five parameters: $\gamma_2$, $\delta_{1,2}$, and $J_{1,2}$, we impose
a fixed value of the total optical power $I_c=I_{c1} + I_{c2}=100$ kW.
In addition, the smaller detuning $\delta_1$ has to be as big as possible. These assumptions, together with the conditions in Eqs.\,\eqref{eqJ} and \eqref{gammas_ratio}, specify all the parameters uniquely. It is easy to show that, if $\gamma_{1,2}\ll|\delta_{1,2}|$, they are given by:
\begin{equation}\label{approx_params}
 \frac{\gamma_1}{\gamma_2} \approx \frac{1}{4} \,,\;\,
 \delta_1 \approx \sqrt[3]{\frac{J}{4}} \,, \;\,\delta_2 \approx -\sqrt[3]{2J} \,,\;\,
 \frac{ I_{c1}}{I_{c2}} \approx \frac{1}{2}\,.
\end{equation}
With the parameters of the facility, we have $\delta_1=4190\,{s}^{-1}$ and $I_{c1}=33$ kW. The resulting optimized noise spectrum $S_F^{\rm opt} = 2\hbar |\chi^{-1}|+S_F^{\rm add}$
is shown in the first row and left column of Fig.\,\ref{fig:plots}.
As we can see, 100kW of circulating power suffice not only for demonstration of a mechanical test object with $\chi^{-1}<m\Omega^2$, but also to achieve a sub-SQL sensitivity in a relatively broad band. To explore the possibilities, we also show the case with an optical loss of $A^2\sim10^{-5}$ per bounce in the left column of Fig.\,\ref{fig:plots}.

\subsection{Large-scale interferometers}
\label{subsec:ligo}

For future large-scale GW detectors, two strategies of implementation of the negative inertia are possible. The first one is just the scaled up version of the setup considered in this paper, which includes three optical pumpings: two for creating optical springs and the third one used for the measurement. For the first two carriers, the interferometer bandwidth has to be as small as possible: $\gamma_1\approx\gamma_2/4 \approx \gamma_{\rm loss}$. It has to be noted, that taking into account kilometer-scale arms lengths of large-scale GW detectors, $\gamma_{\rm loss}$ can be as small as $\sim1\,{\rm s}^{-1}$.  In the second row of Fig.\,\ref{fig:plots}, we show the resulting curve for the total noise spectral density. It shows significant improvements at low frequencies; however, to really achieve such a sensitivity, the third carrier has to have a bandwidth that is comparable to the GW signals, which is much larger than the bandwidth assumed for the first two carriers. We therefore require additional degrees of freedom to achieve a high bandwidth for the third light, which is only possible if the arm cavities are detuned.

The second and probably more promising strategy is to use only two carriers for both creation of the optical springs and for the measurement. For at least one of these pumpings, the corresponding bandwidth has to be of the same order of magnitude as the signal frequency, $\gamma \sim \Omega$. In this case, the optimization procedure has to take into account both the dynamical and the noise properties of the system, and has to provide the parameters set which not just maximize the mechanical susceptibility $\chi$, but minimize the signal-to-noise ratio of the system. This task will be subject of our next paper.
\section{Conclusions}\label{sec:conclusion}

We have shown that the negative inertia effect is capable of reducing the effective inertia of the test mass and thus can significantly enhance the mechanical response of the interferometer to the GW signals, surpassing the free-mass SQL, over a broad frequency band.  This way of beating the free-mass SQL can be understood from the classical point of view: a better sensitivity is achieved because the test mass now has a much higher response to the force that we would like to measure, and the displacement of the test mass (induced by the external force) is read out without any help of quantum correlations. However, it has to be notice that such an enhancement cannot be achieved by the classical feedback control which can also modify the test-mass dynamics. This is because a classical feedback destroy the quantum coherence, which is maintained in the double optical spring scheme. Such an negative-inertia effect, if demonstrated experimentally, will shed light on the potential of implement quantum feedback for improving the sensitivity of future GW detectors.

\section{Acknowledgements}
We thank all our colleagues in the LIGO Macroscopic-Quantum-Mechanics (MQM) group for fruitful discussions. F.K.'s and S.D.'s research have been supported by the Russian Foundation for Basic Research Grant No. 08-02-00580-a. S.D., H.M.-E., and Y.C. are supported by the Alexander von Humboldt Foundation's Sofja Kovalevskaja Programme, NSF grants PHY-0653653 and PHY-0601459, as well as the David and Barbara Groce startup fund at Caltech. H.M. and C.Z. have been supported by the Australian Research Council.

\end{document}